\documentclass[aps,twocolumn,superscriptaddress]{revtex4-1}
\usepackage{graphicx,hyperref,epstopdf,amsfonts,amsmath}
\usepackage{amssymb}
\usepackage{upgreek}
\usepackage{color}
\usepackage{epstopdf}

\begin{document}
	
	\title{Elasticity-based polymer sorting in active fluids: A Brownian dynamics study}
	
	\author{Jaeoh Shin}
	\affiliation{Max Planck Institute for the Physics of Complex Systems, 01187 Dresden, Germany}
		
	\author{Andrey G. Cherstvy}
	\affiliation{Institute for Physics and Astronomy, University of Potsdam, 14476 Potsdam-Golm, Germany}	
	\author{Won Kyu Kim}
	\affiliation{Institut f\"{u}r Weiche Materie and Funktionale Materialen, Helmholtz-Zentrum Berlin,  14109 Berlin, Germany}
	
	\author{Vasily Zaburdaev}
	\affiliation{Max Planck Institute for the Physics of Complex Systems, 01187 Dresden, Germany}
	
	\begin{abstract}
		{While the dynamics of polymer chains in equilibrium media is well understood by now, the polymer dynamics in active non-equilibrium environments can be very different. Here we study the dynamics of polymers in a viscous medium containing self-propelled particles in two dimensions by using Brownian dynamics simulations. We find that the polymer center of mass exhibits a superdiffusive motion at short to intermediate times and the motion turns normal at long times, but with a greatly enhanced diffusivity. Interestingly, the long time diffusivity shows a non-monotonic behavior as a function of the chain length and stiffness. We analyze how the polymer conformation and the accumulation of the self-propelled particles, and therefore the directed motion of the polymer, are correlated. At the point of maximal polymer diffusivity, the polymer has preferentially bent conformations maintained by the balance between the chain elasticity and the propelling force generated by the active particles. We also consider the barrier crossing dynamics of actively-driven polymers in a double-well potential. The barrier crossing times are demonstrated to have a peculiar non-monotonic dependence, related to that of the diffusivity. This effect can be potentially utilized for sorting of polymers from solutions in \textit{in vitro} experiments.}
	\end{abstract}
	
	\maketitle
\section{Introduction}
\label{sec-introduction}

Active fluids composed of self-propelled particles such as motile bacteria, crawling cells, sperm cells, artificial microswimmers \cite{review13, review12b, abpcrowded16, elgeti-microswimmers, stark16review, review15,  gole07, wink15, leonardo13prl, leonardo13, sperm13, catereview12, kroy16, reich16, berenike10, berenike14, zaburdaev14, crawling93,stark17} are inherently out-of-equilibrium systems. An active particle continuously consumes energy generated via internal mechanisms, external fields, or the reservoir energy, as required for its persistent motion \cite{catereview12, lowen14, review12b, review13}. Actively-driven systems often exhibit peculiar features absent at equilibrium conditions \cite{review12b, cacciuto14pressure, catereview12, abpcrowded16, marconi16,stark17}. As an example, diffusive motion of the tracer particles in a medium consisting of swimming bacteria can be characterized by the mean squared displacement (MSD) that on a certain time interval increases faster than linearly in time \cite{metz14, metz16},
\begin{equation} \text{MSD}(t) \propto t^{\alpha} \end{equation}
with $\alpha > 1$ \cite{lipchaber00}. At long times, the scaling of diffusive motion turns normal ($\alpha=1$), but the diffusivity is $2-3$ orders of magnitude larger than in a passive viscous medium \cite{lipchaber00}. The diffusivity of a spherical tracer in active media can vary non-monotonically with the tracer size \cite{nonmonotonicdiffusion16}.

In general, the systems of self-propelled particle cease to follow the equilibrium thermodynamics Boltzmann distribution \cite{cates09, maggi15, leonardo13prl, lipchaber00, abpcrowded16, marconi16, harder14, leonardo13prl, argun16}. The geometry-dependent pressure created by the active particles \cite{cacciuto14curvature, smallen15} can e.g., cause a spontaneous rotation of micron-sized gears in bacterial medium \cite{leonardo09,ratchetmotor10, ratchetmotor10b}. For deformable or responsive tracers, such as polymer chains \cite{kaiser14, kaiser15, cacciuto14, shin15, solon15} or vesicles \cite{vesicle16}, the interplay between the elasticity and active forces reveals interesting effects. The examples include a facilitation of polymer looping \cite{shin15} and a non-monotonous diffusivity of polymers as a function of the chain length \cite{solon15}. The dynamics of polymers in active fluids is relevant to that of various biopolymers in the cellular environments \cite{weitz08}, where the molecular motors generate non-equilibrium conditions \cite{sriram09,gov14, polymer-active-visco15, sakaue16, gladrow16, chak16, activerouse16, winkler17}. Conversely, the fluctuating dynamics of the polymers can be used to infer the nature of active forces present in the system \cite{weitz08}.

Here we study the dynamics of polymer chains in two dimensions (2D) in the presence of active Brownian particles (ABP) \cite{lowen11} by using Brownian dynamics simulations. The recent studies on swelling, collapse, and looping of actively-driven polymers \cite{kaiser14, cacciuto14, kaiser15, shin15,solon15} served as a starting point for the current investigation, with ``active polymers'' being a perspective research direction \cite{abpcrowded16}. Such a 2D system is more relevant to the \textit{in vitro} experimental setups, rather than to \textit{in vivo} settings, as the former are frequently carried out in quasi-2D setups \cite{lipchaber00}.

The polymer chain in the bath of ABPs is not in equilibrium and unusual behaviors can take place \cite{kaiser14, kaiser15, cacciuto14, shin15, solon15, vesicle16, weitz08,sriram09, gov14, polymer-active-visco15, sakaue16, activerouse16, winkler17, chak16, gladrow16}. In this study we find that due to propelling forces of ABPs, the polymer dynamics is greatly facilitated. In particular, the polymer center of mass (COM) diffusivity shows a non-monotonic behavior as a function of both the chain length $L$ and its bending stiffness $\kappa$. The polymer at maximal diffusivity has preferentially bent conformations, maintained by the balance of chain elasticity and propelling forces of ABPs. We also consider the barrier crossing dynamics of polymers in a double-well potential, finding that the crossing times are non-monotonous with $L$ and $\kappa$, too. These results can potentially be utilized for separating polymers based on their length or stiffness \cite{sortingreview} via using active fluids.

The paper is organized as follows. We introduce the model and simulation methods in Sec. \ref{sec-model}. The main results on the polymer diffusive behavior are presented in Sec. \ref{sec-results} and the barrier crossing dynamics of polymer chains is investigated in Sec. \ref{sec-barrier}. Finally, we summarize and discuss our results in Sec. \ref{sec-conclusions}.
\begin{figure}
	\centering
	\includegraphics[width=6cm]{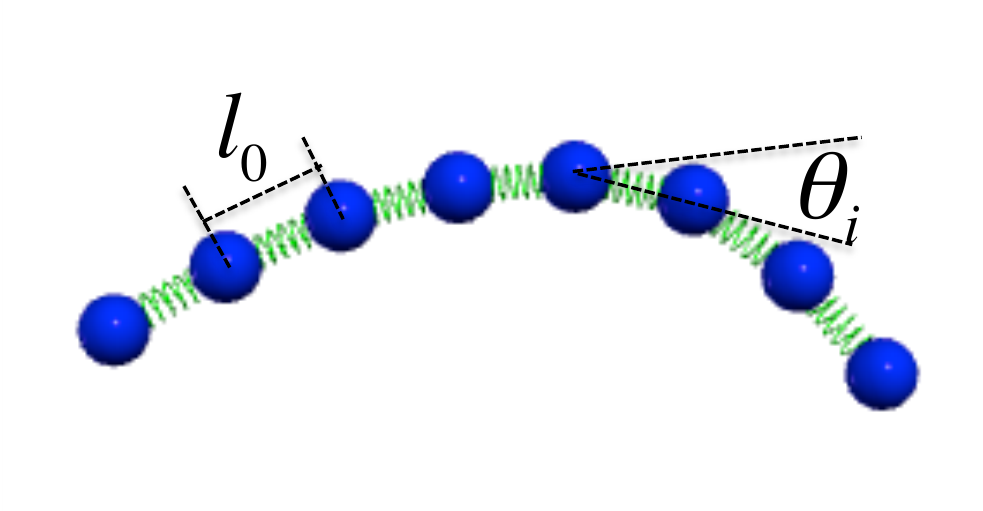}
	\includegraphics[width=6.5cm]{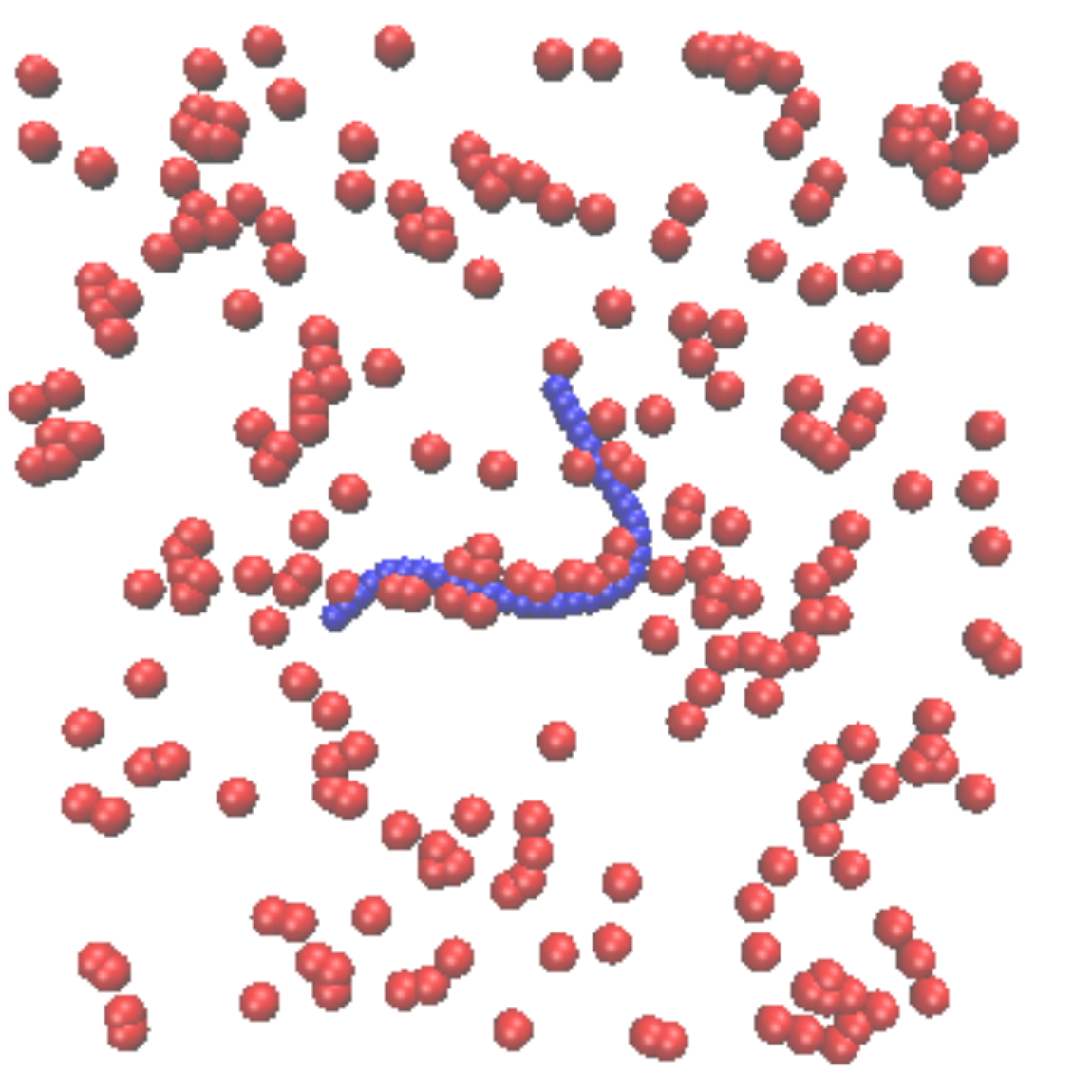}
	\caption{Schematic of the system. (Top) Polymer chain of $n=8$ beads connected by springs. The equilibrium bond length is $l_0$ and the angle between $i$th and ($i+1$)th monomer is $\theta_i$. (Bottom) Configuration of a polymer  (blue chain) in a bath of ABPs  (red spheres). Here $n$=32, $\kappa$=360, and the packing fraction of ABPs is $\phi=0.05$. The figure was rendered using VMD \cite{vmd}. The video files illustrating the chain dynamics are provided in the Supplementary Material.}
	\label{fig-1}
\end{figure}

\section{Model and Methods}
\label{sec-model}

We perform Brownian dynamics simulations of semiflexible polymers in the presence of active particles \cite{review12b} on the plane in 2D.  As a representation of active particles, we use a model of self-propelled ABPs, which can be man-made and used in experiments \cite{lowen11, abpcrowded16}. The position of the $i$th ABP at time $t$ is described by the overdamped Langevin equation \cite{lowen11, abpcrowded16} 

\begin{eqnarray} 
\frac{d \mathbf{r}_{i}(t)}{dt}= -\mu \boldsymbol\nabla U (\mathbf{r}_{i}(t))+ v_{a}\mathbf{n}_i(\psi,t)+ \sqrt{2D_{t}}~ \boldsymbol{\xi}_i(t),
\label{eq-lang-active} 
\end{eqnarray}
where $\mu$ is the particle mobility, $\boldsymbol\nabla  \equiv  \hat{x}\frac{\partial }{\partial x}+\hat{y}\frac{\partial }{\partial y}$, and $\boldsymbol{\xi}_{i}(t)$ is the two-dimensional Gaussian white noise with unit variance in each dimension, $\langle \boldsymbol{\xi}_{i}(t) \cdot \boldsymbol{\xi}_{i'}(t')\rangle= 2\delta_{i,i'}\delta (t - t') $ and $\delta_{i,i'}$ is the Kronecker symbol. Potential $U$ denotes the interaction potential between different ABPs, ABPs and polymer beads, and ABPs with external potential. We refer the reader to Sec. VII of Ref. \cite{abpcrowded16} for the discussion of underdamped dynamics of active particles. Also, the recent examination of inertia effects in some anomalous diffusion processes is instructive \cite{anna-sci-rep-2016}. 

The particle moves with a constant speed $v_a$ along the direction given by angle $\psi$
\begin{equation}\mathbf{n}_i= \left\{\cos\psi, \sin\psi\right\}. \end{equation} This angle is subjected to rotational diffusion, as described by the rotational Langevin equation, \begin{eqnarray}    \frac{d \psi (t)}{dt} =\sqrt{2D_{r}}\xi_{r} (t),\end{eqnarray} where $D_r$ is the rotational diffusivity and $\xi_{r}$ is the Gaussian white noise with unit variance. The rotational diffusion leads to the decorrelation of particle velocity on the {time scale of $\tau=2/D_r$}. For the case of spherical particles of diameter $\sigma$ the value of $D_r$ is related to its translational diffusivity $D_t$ as \cite{membrane15} \begin{equation} D_r=3D_t/\sigma^2. \end{equation} The  strength of particle propulsion is measured in terms of the P\'{e}clet number $$\text{Pe}=v_a\sigma/D_t.$$ The situation of $v_{a}=0$ corresponds to passive Brownian particle, studied previously in the context of macromolecular crowding in e.g. Refs. \cite{shin14NJP, shin15loop, shin15loop2}. 

The polymer is modeled as the bead-spring chain of $n$ monomers of diameter $\sigma$ connected by harmonic springs with the corresponding potential

\begin{eqnarray}
U_{\text{s}}=\frac{k}{2}\sum_{i=2}^{n} (|\mathbf{r}_{i}-\mathbf{r}_{i-1} \rvert-l_0)^2,
\end{eqnarray}
where $k$ is the spring constant and $l_ {0}$ is the equilibrium bond length. Hereafter, the chain monomers are of the same size as the ABPs, see the Discussion section for the effects of ABP size. We choose the Hook's modulus as $k=10^3 k_\text{B}T/\sigma^2$ and $l_0=\sigma$ to prevent the crossing of ABPs by spring sections. The bending energy of the chain is given by

\begin{eqnarray}
U_{b}= \frac{\kappa}{2}\sum_{i=2}^{n-1} \theta_{i}^2,
\end{eqnarray}
where $\kappa$ is the bending stiffness and $\theta_i$ is the bending angle of the $i$th chain segment, see Fig. \ref{fig-1} (Top). For a given value of $\kappa$, the chain persistence length in two dimensions is then $l_p\simeq 2\kappa l_{0}/(k_{\text{B}}T)$. Note that the polymer behaves as a much softer chain in the presence of ABPs due to the enhanced fluctuations \cite{shin15}.

The effects of self-avoidance between different chain monomers and between ABPs and chain monomers are modeled by the Weeks-Chandler-Andersen (WCA) potential \cite{wca},
\begin{eqnarray}
U_ {\text{WCA}} (r_{i,j}) =4\epsilon[(\sigma/r_{i,j})^{12} - (\sigma/r_{i,j})^{6}] + \epsilon,
\label{eq-lang-chain}
\end{eqnarray}
for $r_{i,j} \leq r_{\text{cut}}$ and $U_{\text{WCA}}(r_{i,j})=0$ for $r_{i,j} > r_{\text{cut}}$, where the cutoff distance is $r_{\text{cut}}=2^{1/6}\sigma$. Here, $r_{i,j}=|\mathbf{r}_i-\mathbf{r}_j|$ is the inter-monomer distance and $\epsilon$ is the interaction strength. This potential corresponds to a polymer chain in a good solvent.
To study the barrier crossing dynamics of the chain, we add an external double-well potential along the $x$-axis,
\begin{eqnarray}
U_{\text{ex}}(x)=\frac{A}{4} x^4-\frac{B}{2}x^2.\label{double-well}
\end{eqnarray}
Here the potential width, the distance from one potential minimum to the barrier, is $\Delta x=\sqrt{B/A}$ and the barrier height is $\Delta U=  B^2/(4A)$. Then, the dynamics of the $j$th chain monomer is governed by the following overdamped Langevin equation

\begin{align}
\begin{split}
\frac{d \mathbf{r}_{j}(t)}{dt} = -\mu \boldsymbol\nabla [U_{\text{s}}+\sum_{i \neq j} U_{\text{WCA}}(r_{i,j})+U_{b}+U_{\text{ex}}] +  \\
\sqrt{2D_{t}}\boldsymbol{\xi}_j(t). \label{eq-lang-passive} 
\end{split}
\end{align}

An important parameter of the medium is the density of ABPs. To fix the density, we use the periodic boundary conditions with the square box of area  $\mathcal{L}^2$. We choose $\mathcal{L}$=60 that is larger than the typical size of the polymer, to prevent any artifacts of boundary conditions. The packing fraction of ABPs is defined as 
$$\phi=N_{\text{A}} A_{\text{A}}/\mathcal{L}^2,$$ 
where $N_{\text{A}}$ is the number of particles and $A_{\text{A}}=\pi (\sigma/2)^2$ is the surface area per ABP, see also Ref. \cite{surya15}. We use hereafter $\phi=0.05$. At low packing fractions $\phi$, the interaction between ABPs can be negligible. At high $\phi$, in contrast, some clustering of ABPs and phase separation phenomena can take place \cite{chaikin13, speck13, winkler14, stark14a, stark14b}, that is however beyond the scope of this paper.

To numerically integrate the equations of motion (\ref{eq-lang-active}) and (\ref{eq-lang-passive}), we implement the stochastic Runge-Kutta algorithm \cite{SRK}. We measure the length, the time, and the energy in units of $\sigma$, $t_0=\sigma^2 / D_t$, and thermal energy $k_{\text{B}}T$, respectively. We set below the model parameters as   $\sigma=l_0=1$, $k = 10^{3}$, $D_r=1$, $D_t=1/3$, and $\epsilon = 1$. The important length scales of the system are the chain length $L$, the persistence length $l_p$, and the persistence {length of the ABPs motion $2 v_a/D_r$}. The main features of our results (shown below) will remain the same if we fix the ratio of those lengths. We use the integration time step $\Delta t=2\times 10^{-4}$, so that in our plots, the simulation time of  $t=1$ corresponds to 5000 iteration steps of the evaluation scheme.  Initially, the system is equilibrated for $\sim 10^6$ steps and typically run up to $\sim10^{9}$ iteration steps.

\section{Polymer dynamics in active fluids}
\label{sec-results}

\subsection{From superdiffusive to normal Brownian motion}

We first consider the diffusive motion of a polymer chain in the presence of ABPs and no external potential, $U_{\text{ex}}=0$. From a long trajectory of the chain generated in simulations, we calculate the time-averaged MSD (tMSD) of the polymer COM \cite{metz14},

\begin{eqnarray}
\overline{\delta_X(\Delta)^2}=\frac{1}{\mathcal{T}-\Delta}\int_{0}^{\mathcal{T}-\Delta} [ X({t'+ \Delta})-X({t'})]^2 dt',
\label{tamsd}
\end{eqnarray}
where $$X(t)=\frac{1}{n}\sum_{i=1}^{n} x_{i}$$ is the $x$-coordinate of the polymer's COM. Here, $\Delta$ is the so-called lag time along the trajectory \cite{metz14}.  Moreover, the tMSD is averaged over an ensemble of $N$ independent traces recorded in simulations,

\begin{eqnarray}
\langle \overline{\delta_X(\Delta)^2} \rangle = \frac{1}{N} \sum_{i=1}^{N} \overline{\delta_X(\Delta)^2_{i}},
\end{eqnarray}
with $N$=40 for most of the findings presented below. The tMSD along the $y$-axis is naturally the same as along the $x$-axis in the absence of an external potential. Note that the ensemble of independent trajectories which is required for a satisfactorily smooth behavior of the tMSD is substantially smaller than that for the ensemble averaged MSD \cite{metz14}. tMSD is therefore frequently used in single-particle tracking experiments, where often not so many but rather long traces are generated/available, see e.g. Refs. \cite{lipchaber00, metz14, metz16}. We find that for our system the tMSD is the same as the ensemble averaged MSD (not shown), which means the system is ergodic \cite{metz14}. We thus use $t$ for the lag time $\Delta$ below, for simplicity of the notation. 

\begin{figure*} 
	\centering
	\includegraphics[width=13cm]{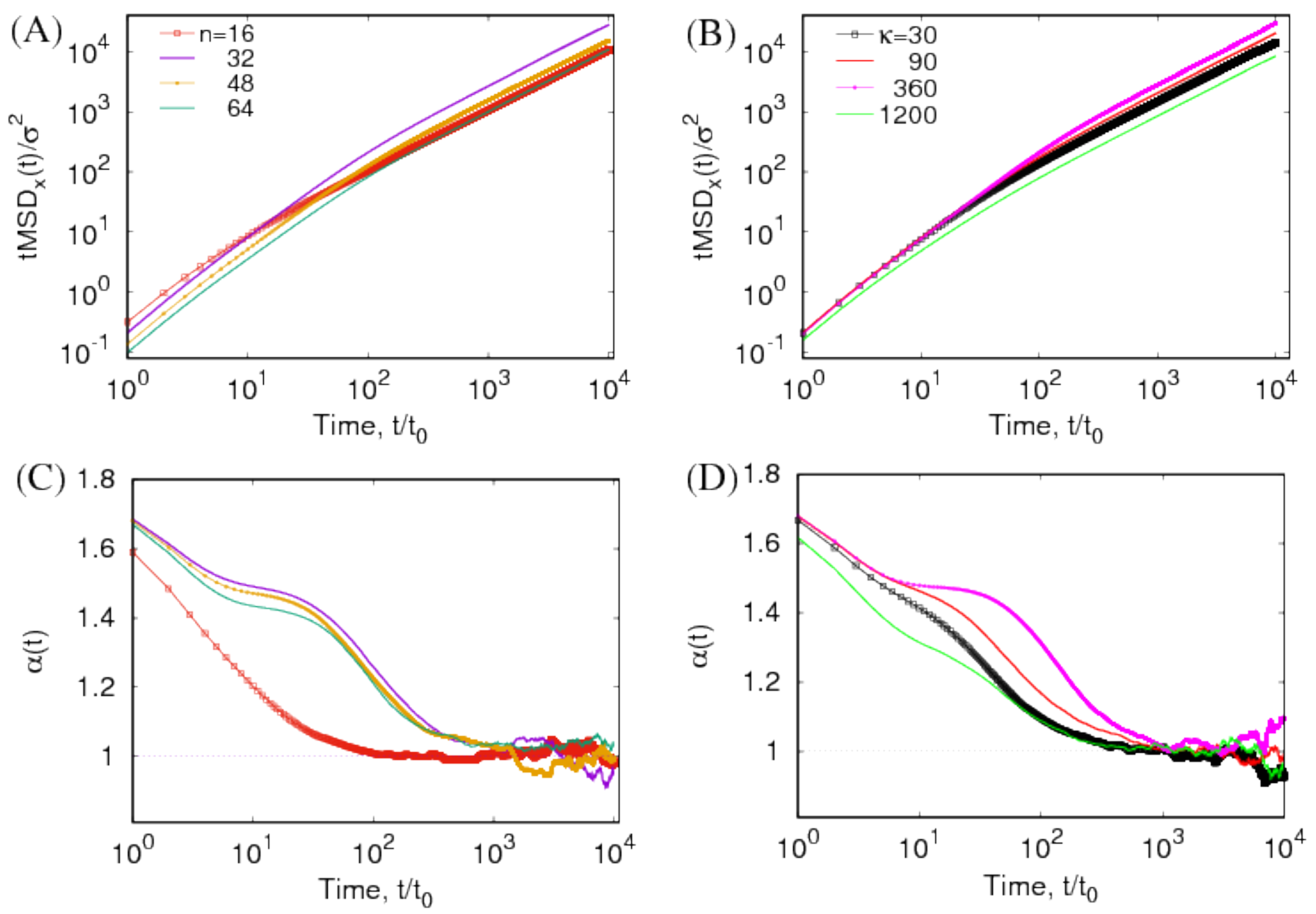}
	\caption{Time-averaged mean squared displacements of the polymer COM and its scaling exponents (\ref{alpha}). (Panel A) tMSD for different polymerization degree $n$ with $\kappa$=200 and (B) tMSD for different bending stiffness $\kappa$ with $n$=32.  (C-D)  Local tMSD scaling exponent for panels (A) and (B). Parameters: the speed of ABPs is $v_a$=10, and the packing fraction of ABPs is $\phi$=0.05.}
	\label{fig-2}
\end{figure*}

We recently demonstrated that the COM motion of a two-dimensional polymer chain is greatly enhanced as the activity of ABPs increases \cite {shin15}. In this study, we focus on how the polymer COM motion depends on the chain length $L=n\sigma$ and its bending stiffness $\kappa$. Active driving by ABPs renders the diffusion of the polymer COM superdiffusive/non-Brownian on intermediate time scales. In Figs. \ref{fig-2}A and \ref{fig-2}B, we show the tMSD of the polymer COM for different chain lengths and bending stiffnesses, respectively.  We compute the time-local scaling exponent of tMSD as \cite{metz14}
\begin{eqnarray}
\alpha(t)=\frac{d [\log{\text{tMSD}(t)}]}{d [\log{t}]}.
\label{alpha}
\end{eqnarray}
We observe in Figs. \ref{fig-2}C and \ref{fig-2}D that the scaling exponent drops from the values $\alpha\approx$1.6-1.8 at relatively short times, to less superdiffusive $\alpha$ values at intermediate times, to finally the normal diffusion behavior with $\alpha=1$ at very long times. Note that here, contrary to our recent study \cite{shin15}, we consider overdamped dynamics for both ABPs and polymer chains and thus one does not expect to see a ballistic regime of the tMSD even in the limit of short times. The noise in $\alpha(t)$ data in the long time limit is due to the worsening statistics as the lag time $\Delta$ becomes comparable to the total time $\mathcal{T}$ in Eq.(\ref{tamsd}), see also Ref. \cite{metz14}.

\begin{figure} 
	\centering
	\includegraphics[width=7.5cm]{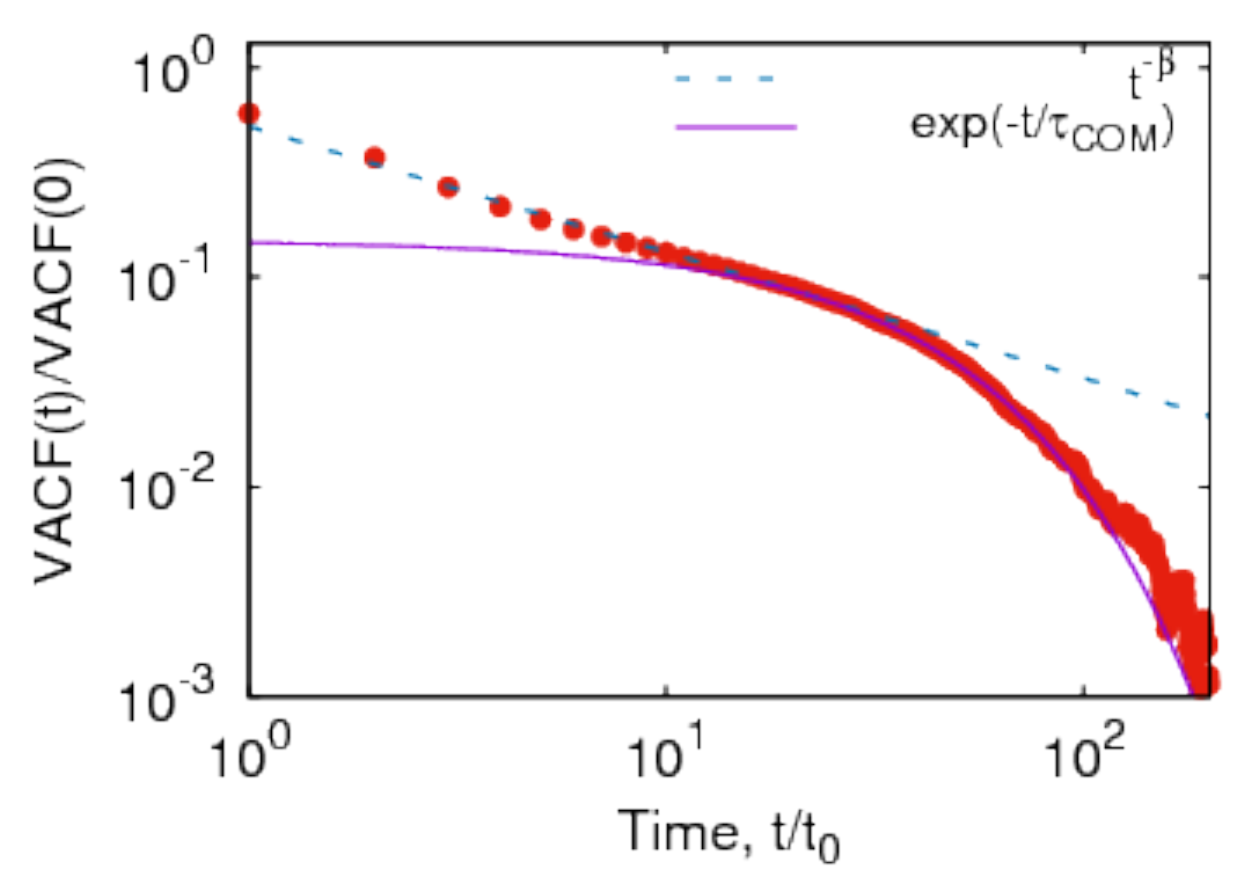}
	\caption{Velocity auto-correlation function VACF($t$) of the polymer COM (red symbols).  Parameters: $n$=32, $\kappa$=360, $v_a$=10, and $\phi$=0.05. In this fitting plot we used $\beta=0.6$ and $\tau_{\text{COM}}=36.5$. }
	\label{fig-3}
\end{figure}

To better characterize the diffusive behavior of the polymer COM, in particular the origin of the superdiffusive behavior at short times, we consider the velocity auto-correlation function (VACF) of the polymer COM, 
$$
\langle V_{\text{COM}}(t)V_{\text{COM}}(0) \rangle,
\label{eq-msd-com-formula}
$$
where $V_{\text{COM}}(t)$ is the COM velocity at time $t$. In Fig. \ref{fig-3}, we show the VACF($t$) for the case of $n$=32 and $\kappa$=360 at which the super-diffusive behavior is most pronounced. The decay of VACF($t$) at short time is close to a power-law, $\sim t ^{-\beta}$, with the exponent $\beta \simeq 0.6$. The power-law decay with the $\beta < 1$ indicates that the tMSD($t$) of the polymer COM, which can be calculated by double integral of the VACF($t$), increases super-linearly,
$$
\text{tMSD}(t)\sim t^{2-\beta},
$$
in the time interval, consistent with our simulation results, see Fig. \ref{fig-2}C and \ref{fig-2}D. For the chains with a less pronounced super-diffusive behavior, the power-law decay regime of the VACF($t$) becomes shorter or disappears and the scaling exponent of tMSD($t$) decreases continuously, see Fig. \ref{fig-2} C and D.
An interesting question is how this power-law decay of VACF($t$) of the polymer COM emerges from the collision of individual ABPs which have exponential decaying correlations \cite{gole07}. This question is, however, out of our scope of this study.

\begin{figure} 
	\centering
	\includegraphics[width=7cm]{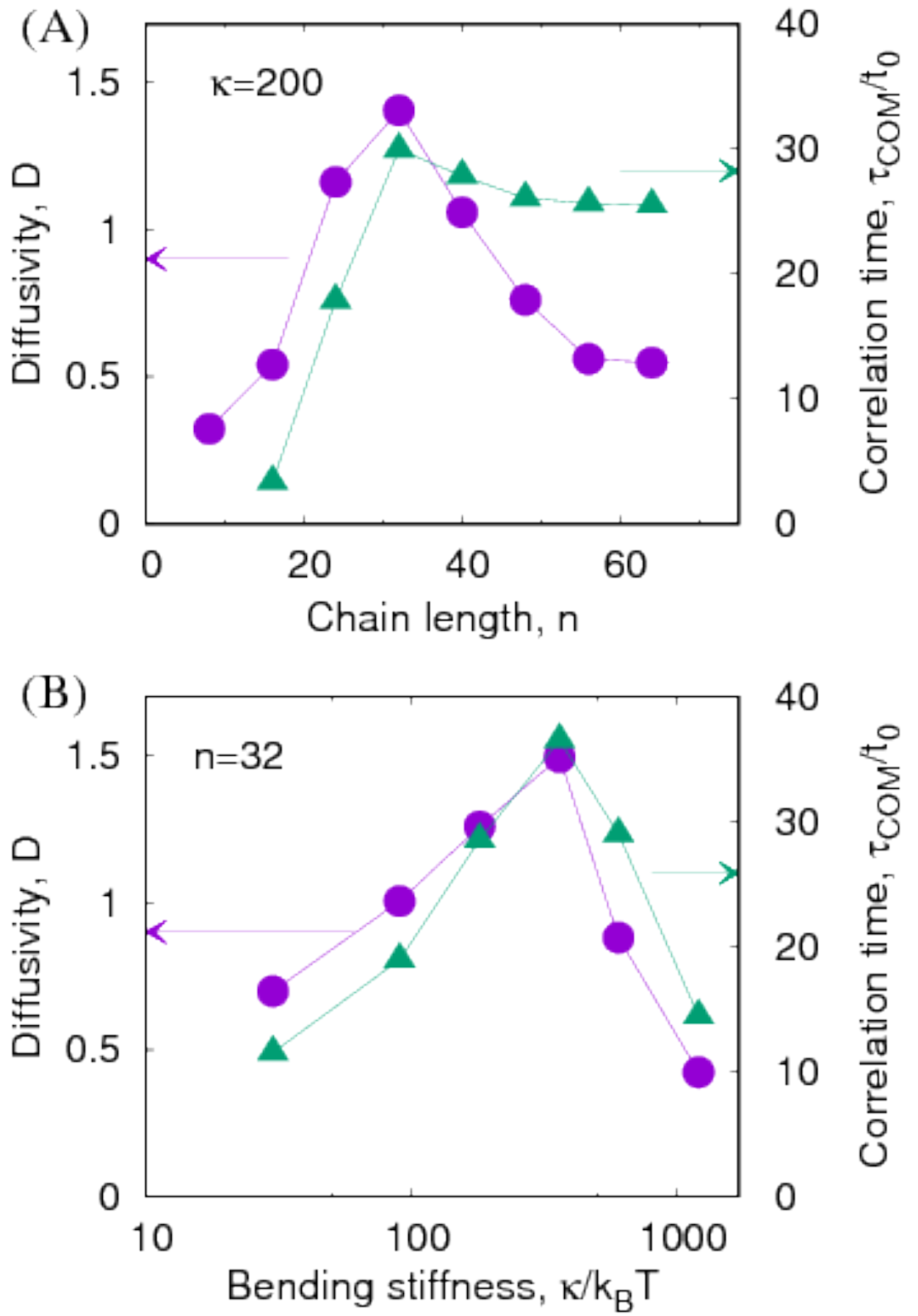}
	\caption{Diffusivity of the polymer COM and the correlation time $\tau_{\text{COM}}$ of VACF: (A) as function of the chain length $n$ with $\kappa=200$ and (B) as function of the bending stiffness $\kappa$ with $n$=32. Other parameters are the same as in Fig. \ref{fig-2}.}
	\label{fig-4}
\end{figure}

At long times, on the other hand, VACF(t) decays exponentially with the correlation time $\tau_{\text{COM}}$, which are shown in Fig. \ref{fig-4} (triangles, the right axis). Physically $\tau_{\text{COM}}$ is the time at which the persistent chain motion start to decorrelate. The correlation time shows a non-monotonic behavior as a function of the chain length $n$ and $\kappa$. The value of $\tau_{\text{COM}}$ is typically, except for the $n=8$ case, much longer than the persistence time of the ABPs, $ \tau=2/D_{r}=2$. In comparison, the VACF of the polymer COM in the absence of ABPs is delta-correlated as is the thermal noise. For the case of ring polymers in two dimensions filled by ABPs \cite{vesicle16}, the VACF is determined by that of the ABPs, independent of the chain elastic properties. The exponential decay of VACF at long times indicates that the tMSD($t$) will increase linearly with time $t$ in this domain, consistent with our simulation results, see Fig.\ref{fig-2}.  

We extract the diffusivity of the polymer COM by fitting the long time limit, in the range $t=[10^3, 10^4]$, of the tMSD($t$) with a linear function. We find that the diffusivity shown in Fig.  \ref{fig-4} (circles, the left axis) also varies non-monotonically with the chain length $n$ and stiffness $\kappa$. The optimal values of $n$* and $\kappa$*, which give rise to the maximum of the diffusivity, are coincident with those of the correlation times. This indicates that the non-monotonic behavior of the diffusivity is resulting from the non-monotonicity of $\tau_{\text{COM}}$. For the case of a fixed chain length (Fig. \ref{fig-4}B), the diffusivity is proportional to the correlation time. This is related to the fact that the diffusivity of ABPs is proportional to the correlation time. However, for the case of varying chain length (Fig. \ref{fig-4}A), the relation is more complicated. The reason is that in this case, not only the correlation time, but also the number of particles pushing the polymer chain, and hence the velocity of the polymer, is changing with the chain length. The non-monotonic dependencies we observe in Figs. \ref{fig-4} are fairly robust with respect to varying $\phi$ and $v_a$, but the optimal chain length and bending stiffness depend on these model parameters.

As can be seen from visualizing the results of simulations at the optimal chain length $n$* or stiffness $\kappa$*, at these parameters the polymer captures the surrounding ABPs for longer time. Therefore, the superdiffusive interval of the polymer COM motion becomes more prolonged, see the $\alpha(t)$ dependencies in Figs. \ref{fig-2}C and \ref{fig-2}D. 
To quantify these observations, we consider various quantities such as end-to-end distance distribution, gyration radii of the polymer, and the Fourier spectrum of chain conformations. As we have shown in Ref. \cite{shin15}, the end-to-end distance distribution and the radii of gyration of polymer significantly change in the presence of ABPs. However, we do not find any distinctive features that indicate the condition of maximum diffusivity of the polymer in these two quantities. Therefore, we present only the Fourier mode analysis of chain conformations in the next subsection.

\subsection{Chain conformations via the Fourier modes}
\label{results-B}

\begin{figure}
	\centering
	\includegraphics[width=7 cm]{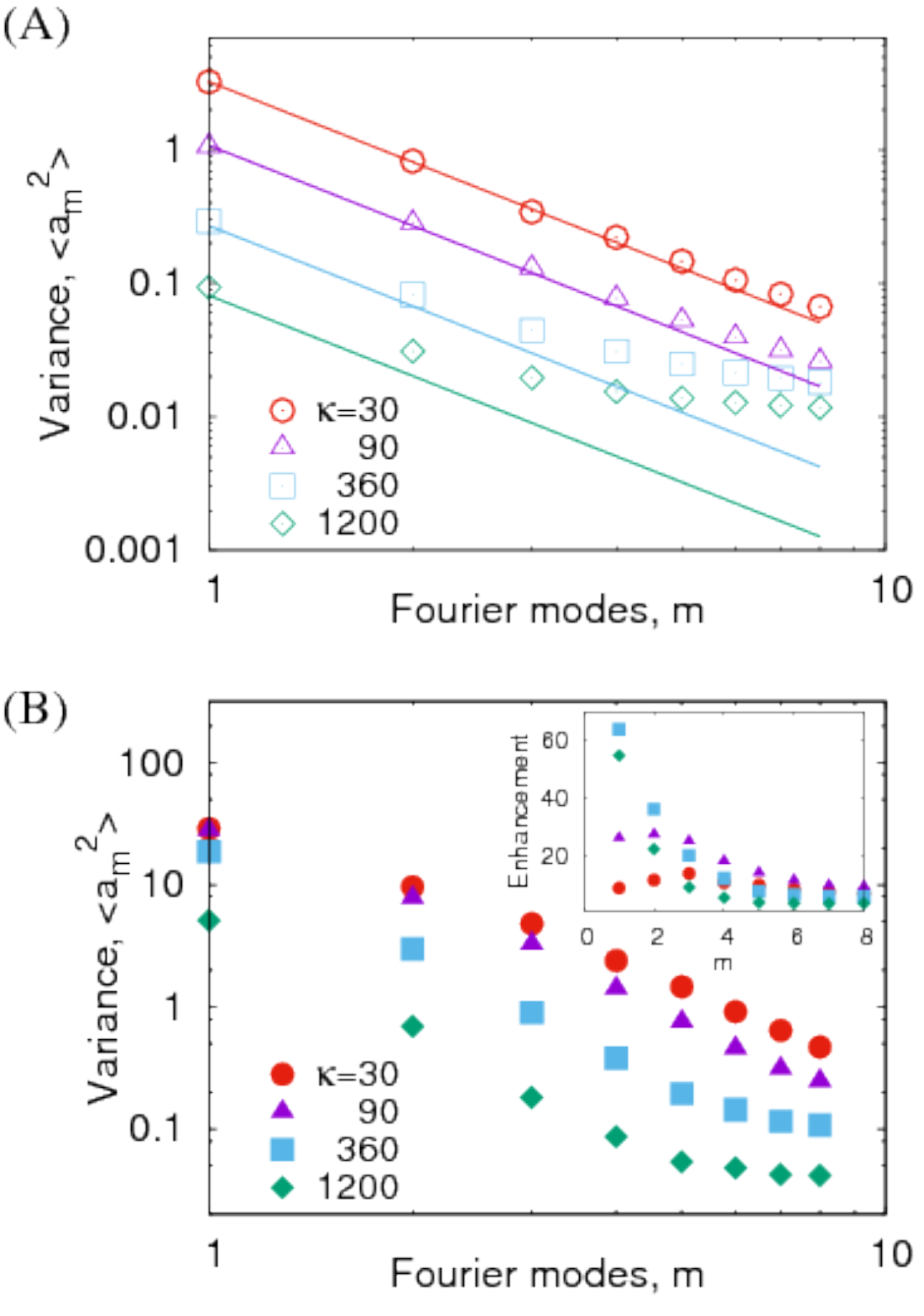}
	\caption{Variance of the Fourier modes $a_m$ for varying chain stiffness $\kappa$. (A) Variance in the absence of ABPs. Theoretical prediction of Eq. (\ref{eq-variance}) is shown as the solid lines. (B) Variance in the presence of ABPs. (Inset) Ratio of the variance in the presence and absence of ABPs. Other parameters are the same as in Fig. \ref{fig-2} B. }
	\label{fig-var}
\end{figure}

We analyze the bending modes of the chain in terms of the Fourier amplitudes of its bending harmonics. The polymer conformations are described in terms of the tangential angle $\theta(s),$ where $s=[0, L]$ is the arc length. Following the method proposed in Ref. \cite{howard93}, the chain conformations are decomposed into the cosine modes,

\begin{eqnarray}
\theta(s)=\sum_{m=0}^{\infty} \theta_{m}(s)=\sqrt{\frac{2}{L}}\sum_{m=0}^{\infty} a_{m}\cos\Big(\frac{m\pi s}{L}\Big). \label{eq-theta-expansion}
\end{eqnarray}
At equilibrium, each mode evolves independently and its variance is given by  \cite{howard93} \begin{eqnarray}
\langle a_m^2\rangle=\frac{k_{{B}}T}{\kappa}\left(\frac{L}{m \pi}\right)^2\propto 1/m^2,
\label{eq-variance}
\end{eqnarray}
in virtue of the equipartition theorem. The mode amplitudes were enumerated from the simulation data via applying the inverse Fourier transform to Eq. (\ref{eq-theta-expansion}), namely \begin{equation}a_{m}=\sqrt{2/L}\int_{0}^{L} ds \theta(s) \cos (m\pi s/L). \label{eq-am-inverse} \end{equation} In what follows we use the discrete approximation of this formula, see Ref. \cite{howard93} for more details.

We first show the variance of the Fourier modes $a_m$ in the absence of ABPs, see empty symbols in Fig. \ref{fig-var}A. The results match well with the theoretical prediction of Eq. (\ref{eq-variance}), shown as the solid lines, for not too stiff chains ($l_p  \leq L$). As the chain becomes stiffer, the simulation results overestimate the theoretical values of $\langle a_m^2\rangle$. The deviation is due to the additional "stretching fluctuations" of the chain in our model so that, as the spring constant $k$ increases, the simulations results (not shown) become closer to the theoretical prediction. However, for the reasons of computational efficiency, we do not use larger spring constants here. 
In Fig. \ref{fig-var}B we show variance of Fourier modes (filled symbols) in the presence of the ABPs. The variances are 1--2 orders of magnitude larger, depending on the mode $m$, compared to that of the chain at equilibrium in the absence of active particles. The enhancement, defined as the ratio of variance in the presence of ABPs to that in the absence of ABPs, is more significant for smaller mode number $m$ as shown in the inset of Fig.  \ref{fig-var}B. We also find that the enhancement of the first two modes ($m$=1 and 2), which determine the large length scale of the polymer conformations, is the largest for the chain of optimal stiffness ($\kappa$=360). This finding indicates that chain conformational fluctuations are highly correlated with the enhanced diffusivity.

The enhancement of fluctuations in actively-driven systems was measured experimentally, among others, for microtubules in the presence of myosin motors \cite{weitz08}. This effect can be interpreted as an increase of the effective temperature in the system, known to facilitate the polymer looping kinetics \cite{shin15}. In contrast to equilibrium systems, here the fluctuations are mainly due to collisions between ABPs and polymer chain, but the energy dissipation occurs via all length scales of the polymer. Thus, it is not surprising that the variance of $a_m$ does not follow the equilibrium scaling relation of Eq. (\ref{eq-variance}). Note also that the distribution of $a_{m}$ amplitudes is always Gaussian in equilibrium systems \cite{reif08}. Physically, this Gaussianity is due to the absence of correlations in the thermal noise. 

\begin{figure}
	\centering
	\includegraphics[width=7 cm]{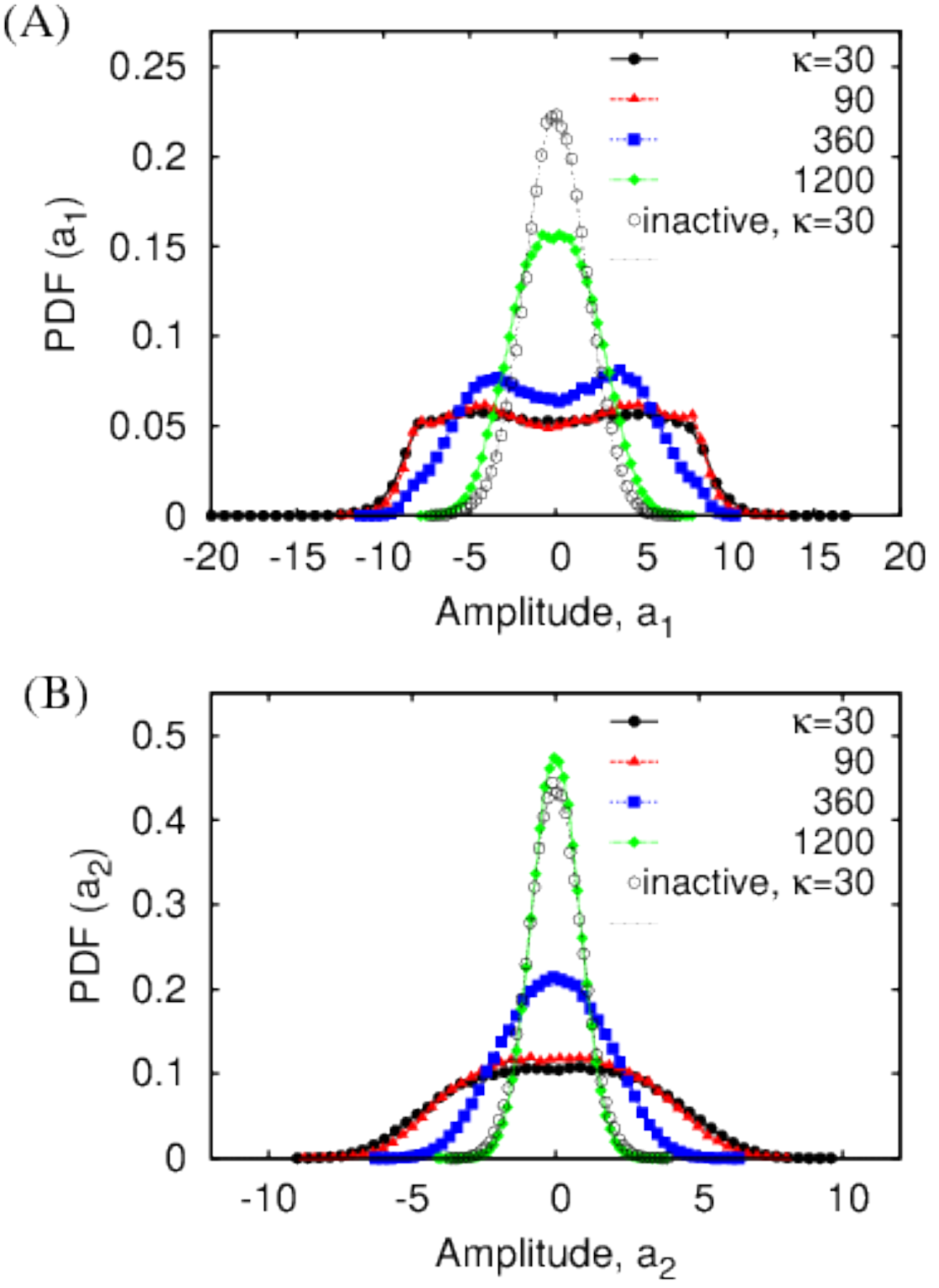}
	\caption{Probability distribution functions (PDF) of the amplitude of the 1st (A) and 2nd (B) Fourier mode. In the absence of ABPs the results are shown as the gray empty circles, together with the theoretical values (\ref{eq-variance}) depicted by the dashed lines. The parameters are the same as in Fig. \ref{fig-var}. }
	\label{fig-hist}
\end{figure}

In the presence of ABPs, however, the distributions of the Fourier amplitudes become strongly non-Gaussian, as we exemplify in Fig. \ref{fig-hist} for the 1st and 2nd mode. We find that for small stiffness parameters $\kappa$ the distribution $p(a_1)$ is roughly uniform in a broad range and for very large $\kappa$ the distributions $p(a_1)$ and $p(a_2)$ reveal a single peak. For intermediate chain stiffness values  $\kappa$, the distribution $p(a_1)$ becomes bimodal, which means the chain adopts preferentially bent conformations. The first Fourier mode corresponds to the half-period of the cosine function and each peak in $p(a_1)$ correspond to the chains in the bent shapes of "$\subset$" and "$\supset$". Such polymer shapes are maintained by the balance between the elastic chain energy and the propelling force of ABPs (compare also to the conformations of actively driven fluid membranes \cite{membrane15}).  This finding is consistent with the mechanism proposed in Ref.\cite{solon15} and our analysis provides a quantitative evidence of the mechanism. For the higher modes (with $m \geq 2$), we find that the distributions are unimodal, but also exhibit non-Gaussian features (the $m=2$ case is shown in Fig.\ref{fig-hist}B)

To summarize, in this section we find that the polymer chains in the presence of ABPs reveal a non-monotonous diffusivity as functions of the chain length $n$ and bending stiffness $\kappa$. At the optimal chain length or stiffness, the polymer has preferentially bent conformations maintained by its elasticity and propelling forces of ABPs. In the next section we show how this effect can be utilized for polymer sorting.

\section{Barrier crossing of polymers in active fluids}
\label{sec-barrier}

Here, we examine the barrier crossing problem for the actively-driven polymers in a double-well potential. The barrier crossing dynamics of polymers in equilibrium media were considered in a number of recent studies \cite{park98, sung01, semi04, sebastian06, shin10}. The crossing times were shown to be strongly dependent on the properties of conformational rearrangement of the polymer in an external potential. In the absence of ABPs, the crossing times of the polymer chains can be rather long, even if the potential barriers are rather low, $\Delta U \leq 1 k_{\text{B}}T$, because all chain monomers need to cross the barrier at the same time. In the presence of ABPs with large $v_a$ values, however, the chain can cross the barrier much shorter time due to the enhanced fluctuations, see Fig. \ref{fig-var}.

Here we consider both the polymers and the ABPs to be subjected to an external potential $U_{\text{ex}}(x)$ acting along the $x$-axis, as expressed in Eq. (\ref{double-well}). Depending on the ratio of the potential width and polymer gyration radius, different barrier crossing scenarios can realize \cite{park98, sung01, semi04, sebastian06, shin10}.
We have chosen the potential width $\Delta X=30\sigma$ is larger than the typical polymer size, so that the polymer COM is placed in either left or right well of the potential. The barrier height also needs to be carefully chosen; if it is too high, the crossing time can be enormously long, but if it is too low, the polymer will move freely in the potential.
	Here we use $\Delta U=6k_{\text{B}}T$ to make the barrier substantial even in the presence of ABPs, but not too high so that a sufficient number of crossing events occurs during our simulation time. In comparison, with the same barrier height but with less active ABPs (smaller $v_a$) or in the absence of ABPs, the barrier crossing events happened extremely rarely (not shown).

As mentioned before, the distribution of ABPs in the external potential $U_{\text{ex}}(x)$ can deviate strongly from the Boltzmann distribution, 
$$
P(x)\sim \exp[-U_{\text{ex}}(x)/(k_{\text{B}}T)],
$$
and the general form of this distribution is not known for a given system. Here, the external force acting on the ABPs is much weaker than the active force, namely
\begin{align}- \frac{dU_{\text{ex}}(x)}{dx} \ll \frac{1}{\mu}v_a,\end{align}
so that the distribution of ABPs is barely affected by the potential. On the other hand, the polymer chains are confined in one of the potential minima.
\begin{figure}
	\centering
		\includegraphics[width=7 cm]{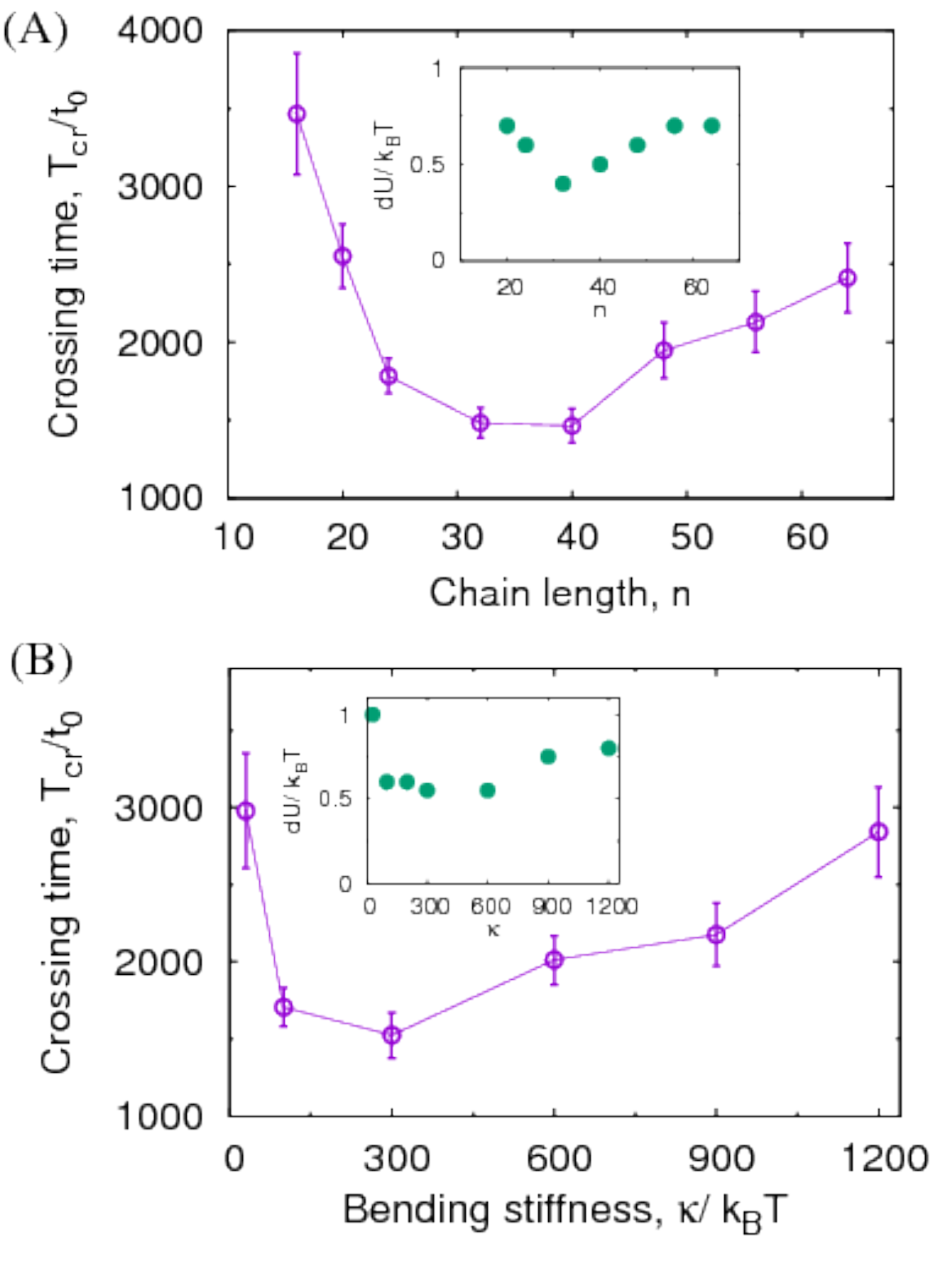}
	\caption{(A) Barrier crossing times $T_{\text{cr}}$ as a function of the chain length $n$, for the bending stiffness of $\kappa$=200. (B) $T_{\text{cr}}$ as a function of $\kappa$, for the chain length of $n$=32. The error bars representing the standard deviation are calculated based on $\sim10^2$ barrier crossing events.  (Insets) Effective barrier heights $dU$, see the main text for details. Here, other parameters are $v_a$=10 and $\phi$=0.05.  }
	\label{fig-crossing-rate}
\end{figure}

We track the COM coordinate $X(t)$ of the polymer which shows a hopping dynamics between the two minima of the potential. We define the crossing time, $T_{\text{cr}}$, as the mean first passage time of the polymer COM from one potential minimum to the other. Figure \ref{fig-crossing-rate} shows the dependence of $T_{\text{cr}}$ on the chain length $n\sigma$ and bending stiffness $\kappa$. The crossing time shows a minimum both at a certain polymerization degree and bending stiffness of the polymer. 

Previously, it was shown that the polymer barrier crossing time in equilibrium can be a non-monotonic function of the polymer length or bending stiffness \cite{park98, sung01, sebastian06, shin10}. In those studies, the polymer barrier crossing dynamics was mapped onto a one-dimensional barrier crossing processes by considering only the COM coordinate and the remaining degrees of the freedom were taken into account by the effective free energy (also known as a potential of the mean force) of the polymer COM. The barrier height of the free energy can be a non-monotonic function of the chain length or stiffness \cite{park98, sung01, sebastian06, shin10}, due to the chain conformational changes, which is the reason of non-monotonic behaviors of the crossing time. Following the approach of Ref. \cite{shin10}, we obtain numerically the effective free energy by using the Boltzmann inversion of the distribution function $P_{\text{COM}}(X)$ along the $x$-coordinate, namely
\begin{align}
F(X)=-k_{\text{B}}T\ln(P_{\text{COM}}(X)).
\end{align}
The effective free energy also exhibits a double-well potential (not shown) and the barrier height of the potential $dU$ is shown in the insets of Fig.\ref{fig-crossing-rate}. The effective barrier height is much smaller than the real barrier height $\Delta U$ and shows a non-monotonic behavior as a function of $n$ or $\kappa$. Following the Kramers' barrier crossing theory \cite{kramers}, we assume that the crossing time of the polymer COM scales as 
\begin{align}
T_{\text{cr}}\sim \frac{1}{D}\exp[dU/(k_{\text{B}}T)]
\end{align}
with the \textit{effective} diffusivity $D$ of the polymer COM and the \textit{effective} barrier height $dU$. The non-monotonous behavior of both $dU$ and $D$ gives rise in a dramatic non-monotonous behavior of the crossing time. In comparison, in equilibrium only the effective barrier height $dU$ can show a nonmonotonous behavior, see Refs. \cite{park98, sung01, sebastian06, shin10}.

\section{Discussion and Summary}
\label{sec-conclusions}

Active fluids are inherently out of equilibrium those are of relevance for a number of living systems. The physical understanding of those systems is, however, far from being complete \cite {review15, abpcrowded16}. Even some basic properties, for example, the distribution of the active particles in external potentials $U_{\text{ex}}$ is only known in some simple setups \cite{cates09, maggi15}. 

Here we numerically studied the dynamics of the actively-driven semiflexible polymers in two dimensions. We found that the ABPs are accumulated in the concave regions of the chain, which results in superdiffusive motion of its COM at intermediate times. At long times, the diffusive motion of the polymer COM becomes normal, but with the diffusivity which was much higher than for the motion without active driving. The diffusivity revealed a maximum versus the polymer length and bending stiffness. The chain at the optimal length or optimal stiffness had preferentially bent conformations, as we have demonstrated examining the chain conformations in the Fourier modes. This occurs when the polymer elastic force and the propelling forces of the ABPs are balanced. 

As an application of the nontrivial behavior of the polymer diffusion in active fluids, we also considered the polymer dynamics with the ABPs in the presence of a double-well potential. We found that, as the activity of ABPs increases, the crossing time is shown to be greatly decreased. The crossing time also showed a non-monotonous dependence with the chain length $n$ and bending stiffness $\kappa$. This is because of a non-monotonous behavior of the diffusivity and the effective barrier height versus $n$ and $\kappa$. This suggests that the polymer chains can be separated from mixtures based on their length or bending stiffness \cite{solon15}, those are important for a number of practical applications \cite{sortingreview}. This scenario might be possible experimentally, for example for sorting of stiff biopolymers such as microtubules or actin filaments immersed in a fluid of active colloidal particles or bacteria \cite{abpcrowded16} in combination with microfabrication channel \cite{leonardo13}. In experiments, it will be important to choose proper system parameters; the activity of the fluids and the potential barrier should be comparable. The former can be controlled, for example by varying the energy source in the fluids \cite{lipchaber00, gole07}, and the latter, by designing the geometry of the channel \cite{leonardo13}.

Note that in our study we considered a single chain in a simulation box, hence the polymer-polymer interactions were absent. For a mixture of many chains, the polymer-polymer interactions could change the crossing dynamics. However, if the polymer density is not very high, the effects of interactions should be minor, not affecting our main findings and trends. Our simple setup with the close-contact potentials neglects also the long-ranged hydrodynamics interactions \cite{gomp09, review13, abpcrowded16}. The latter can govern, among others, some energy transfer reactions and tune collective effects in actively-driven systems, such as those in a 2D diffusion of micron-sized spheres driven by swimming bacteria  \cite{lipchaber00}. Nevertheless, we expect the main features of our findings to stay valid in real systems, particularly when the hydrodynamic effects can be accounted for via a renormalized friction coefficient.

Finally, we have considered that the ABPs are of the same size as the chain monomers. For the case of bigger ABPs, we found that the "capturing" of ABPs by the polymer is  not possible and the non-monotonous behavior of the diffusivity of the polymer COM disappears (results not shown). For smaller ABPs, since the rotational diffusivity scales with the diameter of the particle as $D_r\sim \sigma^{-3}$, the persistence length of the ABPs' motion decreases very rapidly as $\sim \sigma^{3}$. In this case, the distribution of ABPs can be mapped onto the Boltzmann-like distribution, but with a higher "effective" temperature \cite{cates09}, and the non-monotonous diffusive behavior of the polymer chain will disappear. The typical size of self-propelling colloidal particles is in the range of 0.1--10 $\mu$m and the length of the biopolymers such as microtubules can be up to 10 $\mu$m long. Therefore, it would be possible to choose the proper experimental parameters that could demonstrate the validity of our main findings experimentally.  

\section*{Acknowledgements}
We acknowledge S. Saha and S. K. Ghosh for useful discussions.

\end{document}